# Photoinduced Temperature Gradients in Sub-wavelength Plasmonic Structures: The Thermoplasmonics of Nanocones

*Joao Cunha\*, Tian-Long Guo, Alemayehu Nana Koya, Andrea Toma, Mirko Prato, Giuseppe Della Valle, Alessandro Alabastri and Remo Proietti Zaccaria\**

Joao Cunha, Dr. Tian-Long Guo, Prof. Remo Proietti Zaccaria
Cixi Institute of Biomedical Engineering,
Ningbo Institute of Materials Technology and Engineering,
Chinese Academy of Sciences,
ZhongGuan West Road 1219,
Ningbo, 315201,
China
E-mails: cunha.joao@nimte.ac.cn, remo.proietti@iit.it, remo.proietti@nimte.ac.cn

Joao Cunha
University of Chinese Academy of Sciences,
19 A Yuquan Road,
Beijing, 100049,
China

Prof. Giuseppe Della Valle
Dipartimento di Fisica Politecnico di Milano
and
Istituto di Fotonica e Nanotecnologie
Consiglio Nazionale delle Ricerche
Piazza Leonardo da Vinci, 32
I-20133 Milano
Italy

Prof. Alessandro Alabastri
Department of Electrical and Computer Engineering,
Rice University,
Houston 77005, Texas,
United States of America

Dr. Alemayehu Nana Koya, Dr. Andrea Toma, Dr. Mirko Prato, Prof. Remo Proietti Zaccaria
Istituto Italiano di Tecnologia,
Via Morego 30, Genova 16163, Italy

**Abstract.** Plasmonic structures are renowned for their capability to efficiently convert light into heat at the nanoscale. However, despite the possibility to generate deep sub-wavelength electromagnetic hot spots, the formation of extremely localized thermal hot spots is an open challenge of research, simply because of the diffusive spread of heat along the whole metallic nanostructure.

Here we tackle this challenge by exploiting single gold nanocones. We theoretically show how these structures can indeed realize extremely high temperature gradients within the metal, leading to deep sub-wavelength thermal hot spots, owing to their capability of concentrating light at the apex under resonant conditions even under continuous wave illumination. A three-dimensional Finite Element Method model is employed to study the electromagnetic field in the structure and subsequent thermoplasmonic behaviour, in terms of the three-dimensional temperature distribution. We show how the latter is affected by nanocone size, shape, and composition of the surrounding environment. Finally, we anticipate the use of photoinduced temperature gradients in nanocones for applications in optofluidics and thermoelectrics or for thermally induced nanofabrication.

1. Introduction

When electromagnetic (EM) radiation is incident upon a plasmonic nanostructure, it is either absorbed by the nanostructure or scattered back to the surrounding environment. A substantial portion of the absorbed EM energy eventually becomes heat owing to the thermoplasmonic property of the nanostructure [1]. This property, that depends on size, shape and material composition of the nanostructure, defines its capability to convert EM radiation into heat. The generated heat is then dissipated, first in the nanostructure and, on a longer time scale, into the surrounding environment, turning the plasmonic nanostructure into a nanosource of heat [1] and enabling practical applications such as photothermal cancer treatment [2], thermally driven chemical synthesis [3], photo-catalysis and water splitting [4], solar assisted steam generation [5] and micro-and nano-optofluidics [6, 7].

These applications often employ continuous wave (CW) illumination [2–11] which presents experimental and theoretical simplifications with respect to pulsed illumination [12]. Experimentally, solar illumination [5] or CW lasers are typically employed [9–11] whereas, from a theoretical point of view, the CW illumination simplifies the calculation of the temperature distribution through the use of a steady-state analysis [13, 14]. In this respect, the actual steady-state temperature distribution in the nanostructure and in the surrounding environment is governed by the balance between the absorbed incoming EM energy and the outgoing heat transfer to the environment [13].

The size, shape and material composition of the plasmonic nanostructures strongly influence the heat generation which can be highly non-uniform and spatially localized at the nanoscale [8, 13–16]. One of the proposed ways to generate strong spatially localized heating involves nanostructures with strong localized enhancement of the EM near field, i.e. *EM hotspots* [13]. However, despite the localized heating, as the thermal conductivity of the plasmonic material is typically much higher than the thermal conductivity of the surrounding environment, the steady-state spatial temperature distribution becomes approximately uniform all over the plasmonic nanostructure in a very short time [12, 17]. Therefore, upon CW illumination of single nanostructures, large temperature inhomogeneities featuring spots with strongly localized temperature, i.e. *thermal hotspots*, do not usually occur [14, 17]. Nevertheless, the generation of thermal hotspots under CW illumination can be achieved either by designing arrangements of multiple nanostructures [10, 18–20] or by employing an extremely focused laser beam for localizing the EM field in a small region of an extended (large) nanostructure [17].

Here we propose a different approach aiming to the realization of a non-uniform temperature distribution within single sub-wavelength plasmonic nanostructures under CW illumination by fully exploiting the role of the electromagnetic resonances of the nanostructure and their associated generated heat. Among the large variety of possible nanostructure shapes, metallic cones [21] are emblematic plasmonic nanostructures for concentrating EM energy in a single hotspot localized at their apex [22]. This characteristic has been exploited for example in precise molecular sensing applications through Tip-Enhanced Raman Spectroscopy (TERS) techniques [23–25]. In the present study, this characteristic is in turn employed to generate a strong heat source located at the cone apex under particular EM resonant conditions (i.e., a resonance spatially localized at the apex), hence simplifying the thermoplasmonic configuration with a single dominant heat source (as opposed to distributed heat sources along the cone body and base). Furthermore, owing to the capability of conical structures to confine light and heat in a specific volume (i.e., the apex), we aim at generating sufficiently large temperature gradients for temperature-dependent applications.

As analytically solving the EM scattering problem for a cone is challenging [26], its EM properties have been typically studied through numerical simulations, for example by adopting the Finite-Difference Time Domain (FDTD) method [27–29] or the Finite Element Method (FEM) [30]. Similarly, the associated photothermal effects have been studied both analytically [31–33] and numerically, the latter case addressed by the Boundary Element Method (BEM) [32], FDTD [21] and FEM [34–36]. Some of these modeling studies [21, 33–36] have observed temperature inhomogeneities in conical structures with dimension over the diffraction limit ($\mu$m-long tips), where the inhomogeneities are usually considered negligible to affect the addressed applications [22, 24]. Similarly, recent sensing applications [30, 37] based on arrays of ∼ 100 nm-long nanocones [38–40] have leveraged on the excitation of the characteristic EM hotspots of the nanocones while the study of their associated thermoplasmonic response was not addressed.

In this work we investigate the steady-state thermoplasmonic response of a single free-standing sub-wavelength nanocone embedded in different media under CW illumination through threedimensional (3D) FEM modeling. In order to guarantee the robustness of our model, a validation step was first performed by comparing an analytically solvable case with our 3D FEM numerical results, having taken the well-known case [1, 13, 17] of a gold nanosphere under CW illumination as a benchmark. The validated 3D FEM model was then employed to calculate the EM field distribution, the heat generation density and the corresponding temperature distributions inside and outside gold nanocones with several realistic dimensions under CW illumination at visible and near-infrared wavelengths. In particular, we have investigated the influence of the nanocone geometrical parameters and surrounding environment (thermal conductivity) on the thermoplasmonic properties under an incident plane wave linearly polarized along the nanocone axis. We show how the appearance of a thermal hotspot at the apex is linked with the appearance of an apex EM resonance with a corresponding adequate heat source density, all especially boosted by the conical geometry. Finally, the presence of the highly localized strong thermal hotspot naturally induces a sustained geometry-dependent apex-to-base temperature gradient along nanocones even with sub-wavelength dimensions.

## 2. Model and Geometry

### 2.1. Model Equations

To study the thermoplasmonic response of plasmonic nanocones, we consider a FEM modeling domain encompassing the metallic nanostructure and the surrounding dielectric. The governing equations are numerically solved in the modeling domain using a commercial FEM solver, COMSOL Multiphysics, with appropriate boundary conditions (see **Methods**). The electric field inside and outside the nanostructure is obtained by solving the EM scattering problem [41] in the full vectorial regime of Maxwell equations [42]. Similarly, the position- and frequency-dependent heat generation power density is given as the dissipated EM power density as described by the Joule effect [42]

$$q_{\text{abs}}(\mathbf{r},\omega) = \frac{\varepsilon_0\,\omega}{2}\,\text{Im}\left\{\varepsilon_r(\mathbf{r},\omega)\right\}|\mathbf{E}(\mathbf{r})|^2 \quad (1)$$

where $\varepsilon_0$ is the vacuum permittivity, $\varepsilon_r(\mathbf{r},\omega)$ is the complex relative permittivity of the material located at position **r** and at the angular frequency $\omega$ associated to the incident radiation defined by the total electric field vector **E**. In the following we will consider nanocones of gold embedded in different dielectric media (water, borosilicate glass and silicon nitride). In this situation, we will assume the surrounding dielectric as a non-absorbing medium for the EM field and we will consider the heat exclusively generated by the Joule effect inside the metallic nanocone.

When CW illumination is assumed, the temperature distribution can be calculated by adopting the heat transfer equation in its steady-state form,

$$\nabla \cdot (\kappa \nabla T(\mathbf{r})) = -q_{\text{abs}}(\mathbf{r},\omega) \quad (2)$$

with $\kappa$ being the thermal conductivity and where we have employed the Fourier law $\mathbf{j} = -\kappa \nabla T(\mathbf{r})$ relating the heat flux density **j** to the temperature gradient $\nabla T$. The Fourier law can be used at nanoscale dimensions by incorporating small-size corrections to the thermal conductivity [36].

However, since our estimates show that these corrections would not jeopardize the general conclusions of the present work (see **Methods**), for simplicity they have not been included in the 3D FEM model.

| Material | Relative permittivity $\varepsilon_r$ | Thermal conductivity $\kappa$ (Wm$^{-1}$ K$^{-1}$) |
|---|---|---|
| Water | 1.777 (ref. [17]) | 0.6 (ref. [17]) |
| Gold | ref. [43] | 317.0 (ref. [12]) |
| Glass | 2.310 (ref. [44]) | 4.0 (ref. [44]) |
| Si$_3$N$_4$ | 4.064 (ref. [45]) | 30.0 (ref. [46]) |

Table 1: Material physical parameters used in the 3D FEM model.

The physical parameters describing the materials employed in the present work are presented in **Table 1**. All parameters are assumed to remain constant within the considered temperature ranges. No phase transition [44], bubble formation [47] or mass transfer (such as fluid convection [6, 14]) are considered. The frequency-dependent complex relative permittivity of gold is described by the Lorentz-Drude model following Rakić et al. [43].

## 2.2. Nanocone Geometry

The nanocone is geometrically modeled by a conical shape with a hemispherical cap apex of radius $r$ as represented in **Figure 1**. The hemispherical cap apex shape was chosen to resemble nanofabricated cones where the presence of an approximately spherical protrusion at the apex is commonly found [24, 30, 37, 40, 48]. The geometrical shape of the nanocone is completely defined by three parameters: the cone height $h$, the base radius $d$ and the apex radius $r$.

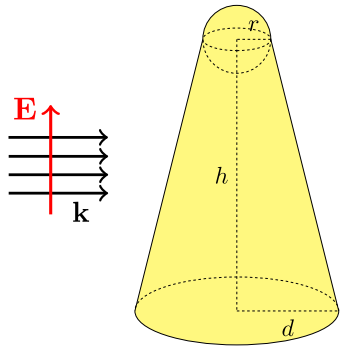

Figure 1: Schematic of the simulated free standing nanocone. The incident illumination representing a wave polarized along the nanocone axis is also depicted.

## 3. Results and Discussion

### 3.1. Thermoplasmonics of Nanospheres

Before tackling the 3D FEM numerical simulation of nanocones, we first consider the simple case of a gold nanosphere of radius $R$ = 50 nm immersed in water [13, 15, 17, 49]. In this case, Equation (2) can be solved analytically [13, 17]. As $\kappa_{gold} \gg \kappa_{water}$, the temperature $T$ in the nanostructure can be approximated as spatially uniform [12, 17]. The steady-state temperature increase is defined as $\Delta T_0 = T - T_0$ where $T_0$ is the initial temperature of the sphere. After calculating the absorption cross-section $\sigma_{abs}$ from the exact analytical result given by Mie theory [41], $\Delta T_0$ is straightforwardly obtained from the analytical expression [1, 13] $\Delta T_0 = \sigma_{abs} I_0/(4\pi \kappa_{water} R)$ where $I_0$ is the illumination intensity impinging on the nanosphere. On the other hand the absorption cross-section can be calculated from the numerical results by $\sigma_{abs} = \int_V q_{abs} dV/I_0$ where the integral of $q_{abs}$ is carried over the nanosphere volume $V$. **Figure S1** summarizes the 3D FEM model and analytical results for the nanosphere showing excellent agreement between numerical results and theory (see **Methods** and **Supporting Information**).

### 3.2. Thermoplasmonics of Nanocones

The validated 3D FEM model was then employed to study metallic nanocones of different dimensions with a meshing strategy that ensures a numerically convergent result (see **Methods** and **Supporting Information**). We illustrate the 3D FEM model results in **Figure 2** for a nanocone with height $h \sim 230$ nm, base radius $d$ = 100 nm and apex radius $r$ = 5 nm immersed in water upon an incoming plane wave with electric field of intensity $I_0$ = 13.33mW/$\mu$m$^2$ polarized along the axis of the nanocone ($z$ axis), **Figure 2(a)**. The choice of water is motivated by the fact that this material is a common surrounding environment in experimental setups and applications [8, 11, 13, 47]. Differently from other theoretical studies employing lower intensities (1mW/$\mu$m$^2$) but similarly to some experimental works [12, 17], here a higher intensity [8, 47] was considered to easily identify thermoplasmonic effects (i.e. photothermal effects resulting in a temperature increase). **Figure 2(b)** shows the resulting electric near field distribution for an incoming wavelength $\lambda$ = 730 nm. The figure depicts an electric field enhancement $|\mathbf{E}/\mathbf{E_0}| \sim 70$ calculated 1 nm away from the nanocone in the surrounding environment. **Figure 2(c)** presents the EM power density, i.e. the heat generation density, with maximum heat generation also located at the nanocone apex. Although EM hotspots and spots of maximum heat localization are typically not spatially coincident [8, 15] as they originate from distinct regions where charge either accumulates or moves freely, respectively, **Figure 2(b)** and **Figure 2(c)** suggest that at particular apex resonances, both the EM hotspot and the generated heat are greatly confined at the apex, thus coexisting in a small volume.

The corresponding temperature distribution is shown in **Figure 2(d)**, where a temperature gradient between cone apex and base is clearly observed, as a result of the presence of a sustained maximum temperature increase localized at the apex, hence demonstrating the possibility of creating thermal hotspots also in sub-wavelength metallic structures.

To illustrate the EM and temperature localization, we present 3D modeling results of the surface charge density and temperature for the visible wavelength range ($\sim$ 400−800 nm), available as a **supporting movie**, for the same nanocone (height $\sim$ 230 nm, base radius = 100 nm and apex radius = 5 nm). The analysis of the 3D surface charge density of complex plasmonic objects such as nanocones or triangular prisms [50, 51] can reveal the spatial origin of the resonances and of the electric near field [21, 50, 52, 53] from the corresponding spatial accumulation (localization) of charge [52, 53]. In a physical picture, upon illumination characterized by an oscillating electric field, the conduction electrons in the nanocone oscillate along the direction of EM polarization inducing a charge accumulation at the nanocone surface associated with spatially localized resonances, e.g. nanocone apex resonance or nanocone base resonance.

In the **supporting movie**, it can be seen that (i) for wavelengths in the 400−600 nm range the surface charge density is asymmetrically distributed on the nanocone surface forming complex patterns associated with the excitation of high-order modes [52, 54] (see **Figure 2(e)** and **Figure**

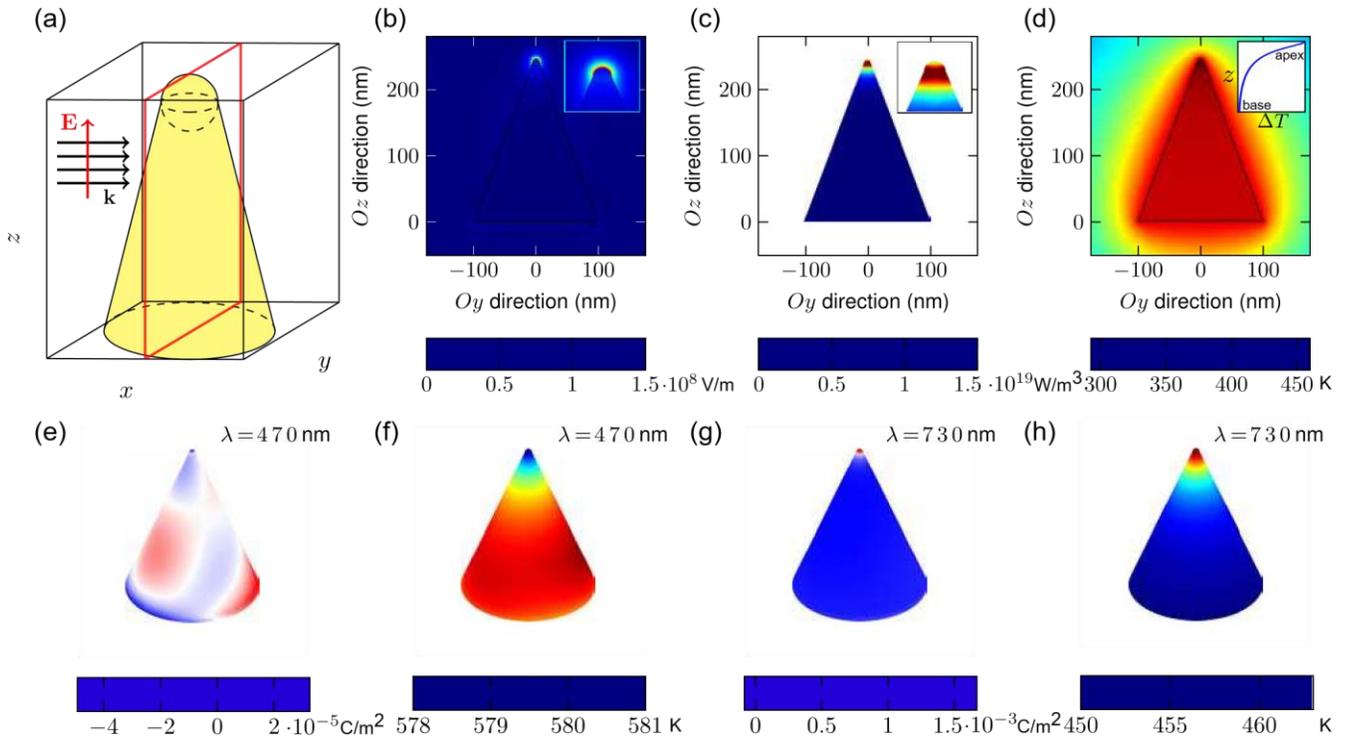

Figure 2: (a) Schematic of the nanocone with the red square representing the cut in the *yOz* plane. The incoming illumination is also shown with a polarization along *z* direction $\mathbf{E} = E_0 \exp(i(k_0 x - \omega t))\mathbf{u}_z$ where $k_0$ is the wavevector. The cone is immersed in water. (b) Electric near field in the cut plane represented in (a) for a nanocone with height $h \sim 230$ nm, base radius $d = 100$ nm and apex radius $r = 5$ nm obtained with incoming $I_0 = 13.33$ mW/µm$^2$ at $\lambda = 730$ nm. The inset represents a zoomed view of 40 nm side. (c) EM power density in the cut plane represented in (a) corresponding to the electric near field in (b). Similarly to (b), the 40 nm side inset represents the zoom view of the apex. (d) Temperature distribution in the cut plane represented in (a) for the heat power density of (c). The inset shows the temperature difference with respect to the base along the axis of the nanocone. The apex-to-base temperature difference $\Delta T$ is 12.8 K. (e) and (f) show 3D views of the surface charge density and the surface temperature distribution at $\lambda = 470$ nm (supporting movie frames). (g) and (h) show 3D views of the surface charge density and the surface temperature distribution at $\lambda = 730$ nm (supporting movie frames).

**S2(a)**) and (ii) a transition to an apex-dominated mode, with charge accumulated at the apex, happens around 600 nm and reaches its maximum accumulation value at $\sim 730$ nm (see **Figure 2(g)**) corresponding to an apex resonance and keeping a similar 3D surface charge density distribution pattern for longer wavelengths. The charge localization at the apex is the result of two combined physical mechanisms: The excitation of localized plasmon polaritons and the lightning-rod effect [23, 55, 56]. Particularly, when an electric field polarized along the nanocone axis is incident with a wavelength allowing for an apex resonance, the excitation of localized plasmons polaritons on the nanocone surface [23, 57] results in rotationally symmetric induced surface charge accumulation with maximum amplitude at the outermost point of the nanocone apex (see e.g. **Figure 2(g)**). In terms of EM field, the charge accumulation at the apex induces an electric near field enhancement (**Figure 2(b)**).

The **supporting movie** also shows the calculated temperature distribution at the surface of the nanocone as function of the incoming excitation wavelength. Interestingly, the temperature distribution in the 400–600 nm range shows a higher absolute temperature in the body than in the apex of the nanocone (see **Figure 2(f)**). This result is easily understood by looking at the surface charge density which, as shown in **Figure 2(e)**, forms complex patterns along the nanocone surface resulting in an EM near field (**Figure S2(a)**) that confirms a distribution of EM power density over the nanocone body, as shown in **Figure S2(b)**. A notable change in temperature distribution takes place around 600 nm, with the temperature maximum and minimum positions being exchanged, establishing a temperature distribution whose maximum is then located at the apex of the nanocone, as highlighted by **Figure 2(h)**. A careful comparison of **Figure 2(f)** and **Figure 2(h)** also reveals a more uniform temperature distribution (and of higher absolute value) in the former case with modest temperature differences along the cone (maximum ∼ 3K). This is once again associated to the EM mode distribution in the cone, with the $\lambda$ = 470 nm case showing a much more extended mode (**Figure 2(e)**) than the strongly localized $\lambda$ = 730 nm case (**Figure 2(g)**).

These results demonstrate that an apex-to-base temperature gradient can be generated in subwavelength metallic nanocone structures under CW illumination. These are intriguing results especially considering that the appearance of temperature gradients within conical structures had been previously observed only in $\mu$m-sized Scanning Tunneling Microscopy tips [58] or Atomic Force Microscopy tips [36]. In turn, our findings show that strong temperature gradients can arise in nanocones with sub-wavelength (< 400 nm-long) dimensions.

To understand the appearance of temperature gradients in illuminated nanocones with subwavelength dimensions we resort to a simple model by studying the thermal resistance of a nanocone assuming the heat source is localized in the spherical volume of its apex, inspired by of height *l* (e.g. *l* = h in **Figure 1**) can thus be calculated through the 1D integration [36] with $R_{\text{th}} = \int_0^l dz/(A_l(z)\,\kappa_{\text{gold}})$ where $A_l$ is the circular cross-sectional area perpendicular to the axis of the truncated nanocone. The result of the integration is plotted in **Figure S3(a)**, showing a strong increase of the thermal resistance when approaching the apex. To further illustrate this aspect, we have divided the truncated cone in three segments of height *l* = h/3 and we have calculated the thermal resistance in each truncated segment using the aforementioned 1D integration model. The integration yields $R_{\text{th,bottom}} < R_{\text{th,middle}} < R_{\text{th,top}}$ with $3.0 \times 10^4$ K/W, $4.4 \times 10^4$ K/W and $8.2 \times 10^4$ K/W, respectively, thus suggesting the establishment of a higher temperature value at the top truncated segment when compared with the temperature in the bottom segments. This temperature profile is confirmed by **Figure S3(b)** where we used the $R_{\text{th}}$ model to estimate the temperature difference $\Delta T$ with respect to the nanocone base along the nanocone axis. Furthermore, **Figure S3(b)** shows a good agreement between the estimated $\Delta T$ and the $\Delta T$ obtained by 3D FEM simulation for a heat source localized in the spherical apex volume (see **Figure S3(b)** inset). These results highlight that the illustrated 1D thermal resistance model can qualitatively explain how a temperature gradient can be established in illuminated nanocones having a dominant heat source localized at the apex.

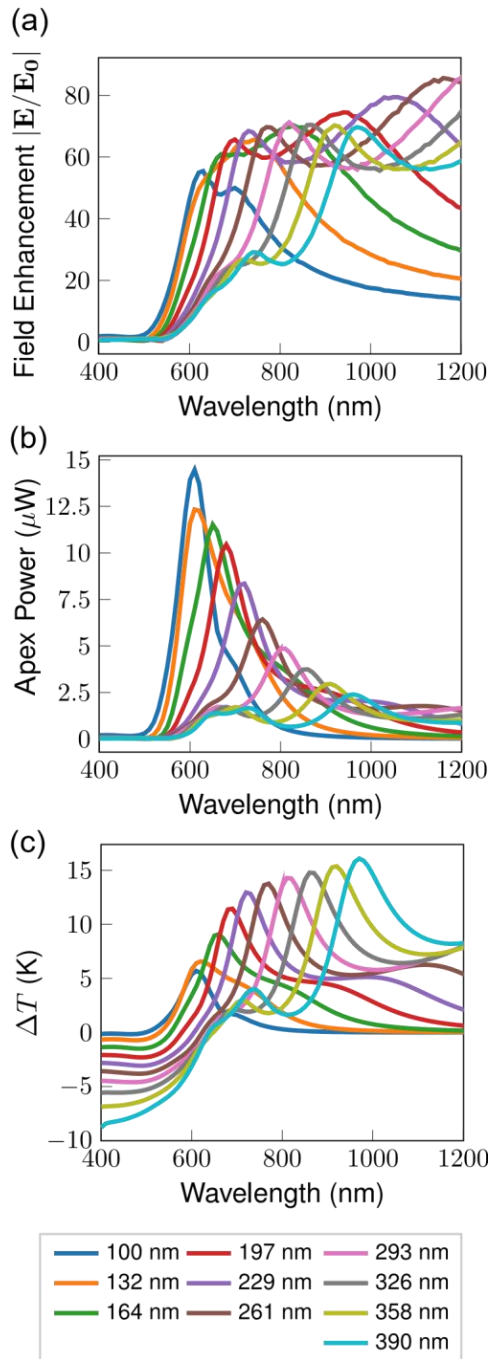

Figure 3: (a) Electric near field enhancement calculated 1 nm away from the outermost point of the nanocone apex, (b) integration of EM power density in the nanocone spherical apex volume (generated heat power at the apex volume) and (c) apex-to-base temperature difference for nanocones of different heights (base radius $d$ = 100 nm and apex radius $r$ = 5 nm) immersed in water with incoming intensity $I_0$ = 13.33mW/$\mu$m$^2$ and electric field polarized along the nanocone axis.

**Figure 2(c)**. In this situation we consider the approximation that the heat transport is mainly one-dimensional (1D) along the cone axis ($Oz$ direction). The thermal resistance $R_{th}$ of a truncated cone

### 3.3. Parameters Affecting the Temperature Gradient

Among the parameters to be considered when it comes to engineer a temperature gradient inside of a sub-wavelength metallic nanocone, we can surely list the cone geometry. In this respect we performed simulations with different nanocone heights ranging from 100 to 390 nm (fixed cone base radius *d* = 100 nm and apex radius *r* = 5 nm) immersed in water, over visible and near-infrared wavelengths with an incoming plane wave with electric field polarized along the nanocone axis. **Figure 3(a-c)** show the near field enhancement |**E**($z$ = apexsurface+1nm)/**E**$_0$|, the calculated EM power at the spherical apex volume (which is a fraction of the total absorbed power that is typically distributed along all the nanocone, as can be obtained from the absorption crosssections reported in Figure S4) and the corresponding apex-to-base temperature difference, i.e., temperature gradient, in the nanocone as function of the incoming wavelength, respectively. For each different nanocone height we can point out (i) a temperature gradient, as promoted by the thermal hotspot, is associated with an apex resonance, EM hotspot, that effectively localizes the EM power in the apex, (ii) while the nanocone first order resonance generally produces the largest near field enhancement [57], the high-order resonances even though carrying less pronounced near field enhancements (more evident in longer cones) tend to more effectively localize the EM power density in the apex (see also **Figure S5**) and (iii) a high field enhancement does not necessarily imply a strong temperature gradient, instead, a strong temperature gradient in the nanocone requires the proper combination of high EM power absorption in the apex volume (thus establishing a dominant heat source localized at the apex) with well defined EM hotspots. Furthermore, by increasing the nanocone height, a red-shift of the apex resonances, evidenced in the electric field enhancement, can be observed [57, 59] together with a red-shift of both the apex power peaks and the temperature gradient. Interestingly, an increase of the cone height also determines an increase of the generated temperature gradient, a trend easily understood by recalling that a longer cone means longer distance between the thermal hotspot, located at the cone apex, and the cone base.

Besides the cone geometry, another important parameter for establishing a strong temperature gradient is the kind of environment surrounding the metallic nanocone. In fact, in the uniform temperature approximation, the temperature distribution does not depend on the thermal conductivity of the surrounding medium up to a constant normalization factor [12, 14, 17]. In contrast, in a nanocone where the heat source is localized at the apex close to the boundary between media with different thermal properties, the temperature distribution depends on the relative ratio of the conductivities of the different media [14]. For this reason we performed EM and thermal 3D FEM simulations with different surrounding materials.

In **Figure 4** we report the temperature distribution, normalized to the apex temperature, for three surrounding environments. Here we chose water (**Figure 4(a)**), followed by borosilicate glass (**Figure 4(b)**) and silicon nitride (**Figure 4(c)**), the latter ones being chosen as two relevant materials from a nanofabrication perspective [44] and having a larger thermal conductivity than water. Our simulation results show that by embedding nanocones in a surrounding environment of larger thermal conductivity (i) the maximum absolute temperature reached at the apex decreases and (ii) the temperature localization at the apex is more pronounced. These results are understood by simply recalling the physical meaning of thermal conductivity as a measure of the ability of a material to conduct heat. In case of surrounding media with sufficiently large thermal conductivity (with respect to the metallic nanocone), the heat produced within the cone will, through a typical diffusive process, more easily find a runaway path to the environment than in the case of media with smaller thermal conductivity, hence effecting the nanocone in terms of lower temperature and higher localization. In other words, when increasing the thermal conductivity of the surrounding environment, more heat will diffuse away through the environment, escaping the nanocone instead of propagating from the tip to the base.

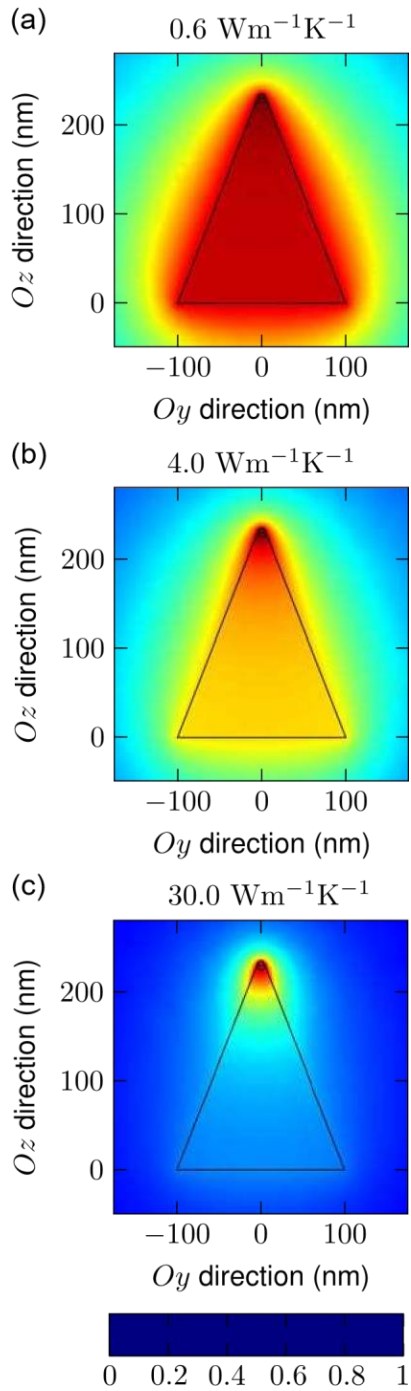

Figure 4: Temperature distributions (normalized to the maximum temperature, reached at the apex) in nanocones of ∼ 230 nm height, 100 nm base radius and 5 nm apex radius embedded under apex resonance excitation. The in (a) water, (b) borosilicate glass and (c) silicon nitride thermal conductivities for each of the materials used in the simulations are highlighted at the top of the figure. The nanocones are excited at their respective apex resonance wavelengths. The absolute temperatures reached at the apex are (a) T ∼ 460 K, (b) T ∼ 329 K and (c) T ∼ 303 K. The apex-to-base temperature differences $\Delta T$ are (a) $\Delta T$ ∼ 12.8 K, (b) $\Delta T$ ∼ 329) $\Delta T$ ∼ 12.1 K and (c) $\Delta T$ ∼ 6.5 K.

Additionally, complementary to the study of the temperature distribution, in **Figure S6** we present the apex-to-base thermal resistance (calculated between the center of the spherical apex and the center of the circular base of the nanocone) of the $h \sim 230$ nm nanocone varying the thermal conductivity of the surrounding environment. Here a 3D FEM simulation mimics an apex resonance with a spherical heat source with power density $1.6 \times 10^{19}$ W/m$^3$ (see **Figure 2(c)**) localized at the nanocone apex as depicted in **Figure S3(b) inset**. The results show that for a weakly conducting surrounding environment (small thermal conductivity), the apexto-base thermal resistance does indeed depend on the dimensions of the cone, with longer cones showing larger thermal resistances. When increasing the thermal conductivity of the surrounding environment, more heat flows towards the environment contributing to reduce the apex-to-base thermal resistance thus lowering the absolute value of temperature increase reached at the apex. For higher values of thermal conductivity of the surrounding environment $\sim 100$ Wm$^{-1}$ K$^{-1}$, the apex-to-base thermal resistances tend to converge to the same value regardless of the cone height and the chosen surrounding environment. Once again, this is easily understood by considering that an increase of the thermal conductivity of the surrounding medium will let the heat escape from the nanocone, hence determining a reduced role of the cone dimension in influencing the thermal resistance.

Finally, an additional aspect affecting the temperature gradient is the small-size effect on the thermal conductivity of gold as we approach the apex. For simplicity, this effect is not incorporated in our model, however we estimate its influence in the **Methods Section** based on an analytical model of the thermal conductivity. The main result is that the value of the thermal conductivity of gold is reduced (with respect to the bulk value) for smaller dimensions and thus, for the same power density delivered to the apex, a comparatively larger apex temperature value and larger apex-to-base temperature difference $\Delta T$ are obtained. This result can be easily understood by employing the 1D thermal resistance model where the thermal resistance (directly proportional to the temperature difference) is inversely proportional to the thermal conductivity of gold and thus, as the thermal conductivity is reduced, the temperature difference would increase. Given the strong reduction in the thermal conductivity for small dimensions (see **Methods Section** and **Supporting Information**), a large enhancement of the gradient towards the apex is expected.

## 4. Conclusion

In this work we have built and validated a 3D FEM model used to reveal the thermoplasmonic properties of a single metallic sub-wavelength nanocone under continuous wave illumination. Our results show that sub-wavelength nanocones can generate a dominant heat source localized at the apex as a result of excitation of an EM hotspot, hence providing the possibility to shrink the spatial dimensions of the heat source from the entire nanostructure down to some of its parts [60]. More importantly, it is also found that the presence of a dominant heat source at the apex can induce a temperature gradient in the nanocone. This finding is especially intriguing considering the small nanocone dimensions (i.e. smaller than the illumination wavelength) as the diffusive nature of heat could hint at the establishment of a uniform temperature over the whole metallic nanostructure given that the thermal conductivity of the nanocone is larger than the surrounding environment [17]. We have then analyzed the role of both the nanocone dimensions and the surrounding environment with respect to the formation of a temperature gradient within the nanocone. We have explained these results through numerical simulations and a 1D thermal resistance analytical model. Overall we have found that through a proper engineering of shape, dimensions and employed materials it is possible to achieve and fully exploit the localization of heat sources within nanostructures [61].

Finally, nanoscale temperature localization and generation of temperature gradients in nanocones are promising features in terms of applications. It is important to stress that to this end we would consider a device integrating a large number of nanocones, a scenario already demonstrated in previous reports [40, 48]. The actual response of the device would be controlled by illumination parameters such as polarization and angle (e.g. tilted illumination) and by the device material and geometrical characteristics such as the spatial arrangement of the nanocones (e.g. randomly or perfectly aligned nanocone arrays). In this regard, we go one step further to anticipate that the effect of tilting the incoming illumination (Figure S7) is of strong importance in assuring an individual response of the nanocone in promoting nanoscale temperature gradients. Furthermore, the electromagnetic and thermal response of the whole device can be strongly dependent on the collective effects of neighboring particles as demonstrated in previous reports [62–64]. Indeed, determining the photothermal response of an array of nanocones is not trivial [18] and requires a dedicated study as an important intermediate step towards applications employing a great number of nanocones. Regarding the potential application areas, it can be envisioned the possibility of nanoscale delivery of heat to small bodies like molecules or cells, to thermally induce the nanosynthesis of materials or to locally activate nanofabrication processes, to enhance optical nanotweezers owing to the presence of high degree of temperature gradient [21] also in conjunction with nanofluidics control [7, 65], or even for thermoelectric power generating devices [66].

## 5. Methods Section

*3D FEM Model:* The 3D FEM modeling study was built and performed in a commercial FEM software (COMSOL Multiphysics).

The EM scattering problem was solved using the "Electromagnetic Wave" (emw) module, where the electric field distribution was calculated through a scattered field formulation using a spherical perfectly matching layer (PML) enclosing the spherical modeling domain (nanocone and surrounding environment) to avoid unphysical reflections on the boundaries. The illumination consisted of a linearly polarized monochromatic plane wave with vacuum wavelength $\lambda$ polarized along the $Oz$ axis. The overall numerical domain geometry is spherical with radius $1.5\lambda$ (size of the modeling domain = $\lambda$ and PML overlayer = $\lambda/2$). The amplitude of the plane wave $|\mathbf{E_0}|$ is calculated for the incoming illumination intensity $I_0$ following $I_0 = n\varepsilon_0 c|\mathbf{E_0}|^2/2$ where $n$ is the refractive index

of the surrounding environment, $c$ is the speed of light and $\varepsilon_0$ is the vacuum permittivity. The refractive index of the surrounding environment is related with the relative permittivity through $n = \sqrt{\varepsilon_r}$. For the calculation of temperature distribution, the "Heat Transfer in Solids" (ht) module was used. The same FEM mesh is employed in both emw and ht modules. The generated heat $q_{abs}$ (Equation (1)) is inserted as a heat source term in the thermal diffusion equation (Equation (2)). For the ht module, spherical infinite elements (IE) enclose the modeling domain, simulating an unbounded background with room temperature $T_0 = 293.15K$, intended to set a temperature $T_0$ very far away from the modeling domain (nanostructure and surroundings). The implemented IE layer was sufficiently thick to ensure convergence of the temperatures inside the modeling domain.

*3D FEM Model Validation:* The 3D FEM model validation results are shown in **Figure S1** where a gold sphere of 50 nm radius immersed in water is illuminated by an incoming plane wave with electric field polarized along $z$ with intensity $I_0 = 1.33\text{mW}/\mu\text{m}^2$. **Figure S1(b)** and **Figure S1(e)** present the electric near field intensity mapping at the nanosphere resonance wavelength, $\lambda \sim 570$ nm. **Figure S1(c)** and **Figure S1(f)** present the heat generation density mapping. **Figure S1(g)** shows the absorption cross-section obtained from the 3D FEM as a function of the wavelength with excellent agreement between the numerical results and the absorption cross-section calculated with Mie theory. **Figure S1(h)** shows the corresponding FEM modeling results for the temperature distribution. The temperature distribution inside the nanosphere is approximately uniform despite the strongly non-uniform heat generation density, as also noted in refs. [8, 14, 17]. In **Figure S1(i)** we show the calculation of the temperature increase $\Delta T_0$ obtained by the 3D FEM model together with the result of the analytical calculation. **Figure S1(i)** shows excellent agreement between the two results thus validating the 3D FEM model.

*Error Analysis:* The main sources of inaccuracies that can affect the 3D FEM model result are inaccuracies in calculation of the electric near fields and inaccuracies in calculations of the temperature distribution. Once a PML and IE overlayers are correctly set (as could be assessed by the validation example, see **Figure S1**), the inaccuracies are linked to meshing errors, reduced number of mesh elements in addition to inappropriate mesh elements for the geometry and the physical processes being modeled. For the physical process modeled, we anticipate an EM hotspot in the vicinity of the apex and a strong heat generation inside the apex volume. Consequently, to resolve these, we appropriately generate a higher mesh density inside the apex volume and implement mesh refinement steps in the apex surface. An example of the generated mesh is shown in **Figure S8**.

To estimate the effect of the size of elements of the FEM mesh in the obtained values of electric-near field and temperature, we perform a typical mesh convergence study by gradually changing the maximum mesh element size with respect to the incoming wavelength $\lambda$ (which sets the simulation domain size) from $0.1\lambda$ to $0.6\lambda$. A smaller maximum element size is generally linked with a larger total number of elements. Along the simulated wavelengths, the total number of elements varies in the range $3-6 \times 10^5$. **Figure S9** shows the result of the mesh convergence study where the normalized near field enhancement 1 nm away from the outermost point of the apex and the normalized temperature of the nanocone apex at the center converge with less than 1% error as the size of the larger elements in the generated mesh decreases (finer mesh) to $0.1\lambda$ — the value used in the 3D FEM model simulations. We note that regardless of the value of the larger element size, the apex surface undergoes mesh refinement, resulting in very fine elements at the surface effectively resolving the EM hotspot (**Figure S8**).

*Temperature and Small-Size Corrections to the Gold Thermal Conductivity:* In the 3D FEM model we considered the bulk values of the thermal conductivity of materials **Table 1** (without small-size corrections). Since the electron mean free path in gold at room temperature is $\sim 40$ nm [67] and the thermal conductivity of gold is mostly contributed by electrons, small-size effects at dimensions smaller than the electron mean free path can potentially influence the thermal conductivity. The thermal conductivity is dependent on the temperature $T$ and on the characteristic dimension of the structure $d_c$, being given by [68] $\kappa_{\text{gold}}(d_c,T) = C_e(T)v_F \ell_{\text{eff}}(d_c,T)/3$ where $C_e$ is the electron heat capacity per unit volume, $v_F$ is the Fermi velocity and $\ell_{\text{eff}}(d_c,T)$ is the size- and temperature-dependent effective electron mean free path. The effective electron mean free path is given by [36] $\ell_{\text{eff}}^{-1}(d_c,T) = \ell_e^{-1}(T) + d_c^{-1}$ where $\ell_e^{-1}(T)$ is the electron mean free path (see supporting information in **Figure S10**) and the dimension $d_c$ can be taken as the diameter of the circular cross-section of the nanocone. The dimension $d_c$ decreases as we approach the apex of the nanocone, yielding a reduction of thermal conductivity $\kappa_{\text{gold}}(d_c,T)$ with respect to the bulk value. In **Figure S10** we present a simple evaluation of the thermal conductivity of gold $\kappa_{\text{gold}}(d_c,T)$ as function of temperature $T$ and dimension $d_c$. While in the temperature range under consideration (300–500 K) the variation in the value of the bulk thermal conductivity of gold is considerably modest, see **Figure S10(a)**, the reduction in the thermal conductivity $\kappa_{\text{gold}}$ due to small-size effects can be dramatic. Indeed, in **Figure S10(b)** $\kappa_{\text{gold}}$ is reduced to ~10% of the bulk conductivity value (taken as in **Table 1**) when $d_c$ = 10 nm (diameter of the nanocone apex). We thus anticipate that the thermal conductivity reduction due to small-size effects in the nanocone apex region would result in a larger absolute temperature value. Correspondingly, the apex-to-base temperature difference would also be expected to increase when the thermal conductivity is reduced, as can simply be analyzed through the inverse relationship between $R_{\text{th}}$ and $\kappa_{\text{gold}}$ in the 1D thermal resistance model. Therefore, we point out that the absolute value of temperature and temperature difference obtained in this study should be taken as a lower bound.

**Supporting Information**

Supporting Information is available from the author. It includes a Supporting movie (.mp4) and a Supporting information PDF file (.pdf) with Figures S1-10. Thermoplasmonics of sphere with $R$ = 50 nm in water. FEM mesh and mesh refinement result. Mesh convergence study. Supporting movie description. Apex-to-base thermal resistance model and results. Thermal conductivity of Au as function of temperature and length scale. Apex-to-base thermal resistance of cones with different heights.

**Acknowledgments**

J.C. acknowledges CAS-TWAS President's Fellowship funding. R.P.Z. acknowledges Ningbo 3315 Innovative Teams Program of the Chinese Academy of Sciences (CAS) (Grant No. Y70001DL01). A.T. and M.P. acknowledge Dr. A. Weber-Bargioni (Molecular Foundry, Lawrence Berkeley National

Laboratory) for helpful discussions and European Union's Horizon 2020 research and innovation programme (SONAR) under the Marie Sk lodowska-Curie grant agreement No. 734690 for funding.

**References**

[1]   G. Baffou, *Thermoplasmonics: Heating metal nanoparticles using light*, Cambridge University Press, **2017**, pp. 1–291.

[2]   Z. Qin, J. C. Bischof, *Chemical Society Reviews* **2012**, *41*, 1191–1217.

[3]   G. Baffou, R. Quidant, *Chemical Society Reviews* **2014**, *43*, 3898–3907.

[4]   A. Naldoni, U. Guler, Z. Wang, M. Marelli, F. Malara, X. Meng, L. V. Besteiro, A. O. Govorov, A. V. Kildishev, A. Boltasseva, V. M. Shalaev, *Advanced Optical Materials* **2017**, *5*, 1601031.

[5]   O. Neumann, A. S. Urban, J. Day, S. Lal, P. Nordlander, N. J. Halas, *ACS Nano* **2013**, *7*, 42–49.

[6]   J. S. Donner, G. Baffou, D. McCloskey, R. Quidant, *ACS Nano* **2011**, *5*, 5457–5462.

[7]   J. C. Ndukaife, A. V. Kildishev, A. G. A. Nnanna, V. M. Shalaev, S. T. Wereley, A. Boltasseva, *Nature Nanotechnology* **2016**, *11*, 53–59.

[8]   G. Baffou, C. Girard, R. Quidant, *Physical Review Letters* **2010**, *104*, 136805.

[9]   J. B. Herzog, M. W. Knight, D. Natelson, *Nano Letters* **2014**, *14*, 499–503.

[10]  Z. J. Coppens, W. Li, D. G. Walker, J. G. Valentine, *Nano Letters* **2013**, *13*, 1023–1028.

[11]  H. Ma, P. Tian, J. Pello, P. M. Bendix, L. B. Oddershede, *Nano Letters* **2014**, *14*, 612–619.

[12]  G. Baffou, H. Rigneault, *Physical Review B - Condensed Matter and Materials Physics* **2011**, *84*, 035415.

[13]  A. O. Govorov, W. Zhang, T. Skeini, H. Richardson, J. Lee, N. A. Kotov, *Nanoscale Research Letters* **2006**, *1*, 84–90.

[14]  G. Baffou, R. Quidant, C. Girard, *Physical Review B - Condensed Matter and Materials Physics* **2010**, *82*, 165424.

[15] G. Baffou, R. Quidant, C. Girard, *Applied Physics Letters* **2009**, *94*, 153109.

[16] A. Alabastri, A. Toma, M. Malerba, F. De Angelis, R. Proietti Zaccaria, *ACS Photonics* **2015**, *2*, 115–120.

[17] G. Baffou, R. Quidant, F. J. García De Abajo, *ACS Nano* **2010**, *4*, 709–716.

[18] G. Baffou, P. Berto, E. Bermúdez Ureña, R. Quidant, S. Monneret, J. Polleux, H. Rigneault, *ACS Nano* **2013**, *7*, 6478–6488.

[19] G. Baffou, E. B. Ureña, P. Berto, S. Monneret, R. Quidant, H. Rigneault, *Nanoscale* **2014**, *6*, 8984–8989.

[20] L. Khosravi Khorashad, L. V. Besteiro, Z. Wang, J. Valentine, A. O. Govorov, *Journal of Physical Chemistry C* **2016**, *120*, 13215–13226.

[21] L. Novotny, R. X. Bian, X. S. Xie, *Physical Review Letters* **1997**, *79*, 645–648.

[22] A. Giugni, B. Torre, A. Toma, M. Francardi, M. Malerba, A. Alabastri, R. Proietti Zaccaria, M. I. Stockman, E. Di Fabrizio, *Nature Nanotechnology* **2013**, *8*, 845–852.

[23] R. P. Zaccaria, F. De Angelis, A. Toma, L. Razzari, A. Alabastri, G. Das, C. Liberale, E. Di Fabrizio, *Optics Letters* **2012**, *37*, 545.

[24] F. De Angelis, G. Das, P. Candeloro, M. Patrini, M. Galli, A. Bek, M. Lazzarino, I. Maksymov, C. Liberale, L. C. Andreani, E. Di Fabrizio, *Nature Nanotechnology* **2010**, *5*, 67–72.

[25] Z. Zhang, S. Sheng, R. Wang, M. Sun, *Analytical Chemistry* **2016**, *88*, 9328–9346.

[26] R. Proietti Zaccaria, A. Alabastri, F. De Angelis, G. Das, C. Liberale, A. Toma, A. Giugni, L. Razzari, M. Malerba, H. B. Sun, E. Di Fabrizio, *Physical Review B - Condensed Matter and Materials Physics* **2012**, *86*, 035410.

[27] F. Demming, J. Jersch, K. Dickmann, P. I. Geshev, *Applied Physics B: Lasers and Optics* **1998**, *66*, 593–598.

[28] R. M. Roth, N. C. Panoiu, M. M. Adams, R. M. Osgood, C. C. Neacsu, M. B. Raschke, *Optics Express* **2006**, *14*, 2921.

[29] A. V. Goncharenko, H. C. Chang, J. K. Wang, *Ultramicroscopy* **2007**, *107*, 151–157.

[30] C. Schäfer, D. A. Gollmer, A. Horrer, J. Fulmes, A. Weber-Bargioni, S. Cabrini, P. J. Schuck, D. P. Kern, M. Fleischer, *Nanoscale* **2013**, *5*, 7861–7866.

[31] P. I. Geshev, S. Klein, K. Dickmann, *Applied Physics B: Lasers and Optics* **2003**, *76*, 313–317.

[32] P. I. Geshev, F. Demming, J. Jersch, K. Dickmann, *Applied Physics B: Lasers and Optics* **2000**, *70*, 91–97.

[33] P. I. Geshev, F. Demming, J. Jersch, K. Dickmann, *Thin Solid Films* **2000**, *368*, 156–162.

[34] A. Downes, D. Salter, A. Elfick, *Journal of Physical Chemistry B* **2006**, *110*, 6692–6698.

[35] X. Chen, X. Wang, *Nanotechnology* **2011**, *22*, 075204.

[36] X. Chen, X. Wang, *Journal of Physical Chemistry C* **2011**, *115*, 22207–22216.

[37] C. Schäfer, D. P. Kern, M. Fleischer, *Lab on a Chip* **2015**, *15*, 1066–1071.

[38] M. Fleischer, C. Stanciu, F. Stade, J. Stadler, K. Braun, A. Heeren, M. Häffner, D. P. Kern, A. J. Meixner, *Applied Physics Letters* **2008**, *93*, 111114.

[39] M. Fleischer, A. Weber-Bargioni, S. Cabrini, D. P. Kern, *Microelectronic Engineering* **2011**, *88*, 2247–2250.

[40] A. Horrer, C. Schäfer, K. Broch, D. A. Gollmer, J. Rogalski, J. Fulmes, D. Zhang, A. J. Meixner, F. Schreiber, D. P. Kern, M. Fleischer, *Small* **2013**, *9*, 3987–3992.

[41] C. Bohren, D. Huffman, *Absorption and Scattering of Light by Small Particles*, Wiley-VCH, **1998**, p. 544.

[42] J. D. Jackson, *Classical electrodynamics*, 3rd, Wiley, New York, **1999**.

[43] A. D. Rakić, A. B. Djurišić, J. M. Elazar, M. L. Majewski, *Applied Optics* **1998**, *37*, 5271.

[44] M. Mutlu, J. H. Kang, S. Raza, D. Schoen, X. Zheng, P. G. Kik, M. L. Brongersma, *Nano Letters* **2018**, *18*, 1699–1706.

[45] A. E. Kaloyeros, F. A. Jové, J. Goff, B. Arkles, *ECS Journal of Solid State Science and*

*Technology* **2017**, *6*, P691–P714.

[46] C. Qiu, Y. Yang, C. Li, Y. Wang, K. Wu, J. Chen, *Scientific Reports* **2017**, *7*, 17046.

[47] G. Baffou, J. Polleux, H. Rigneault, S. Monneret, *Journal of Physical Chemistry C* **2014**, *118*, 4890–4898.

[48] F. Stade, A. Heeren, M. Fleischer, D. P. Kern, *Microelectronic Engineering* **2007**, *84*, 1589–1592.

[49] A. Alabastri, S. Tuccio, A. Giugni, A. Toma, C. Liberale, G. Das, F. D. Angelis, E. D. Fabrizio, R. P. Zaccaria, *Materials* **2013**, *6*, 4879–4910.

[50] J. P. Kottmann, O. J. Martin, D. R. Smith, S. Schultz, *New Journal of Physics* **2000**, *2*, 27. [51] J. P. Kottmann, O. J. Martin, D. R. Smith, S. Schultz, *Optics Express* **2000**, *6*, 213.

[52] S. D'Agostino, F. Della Sala, L. C. Andreani, *Physical Review B - Condensed Matter and Materials Physics* **2013**, *87*, 205413.

[53] R. Marty, G. Baffou, A. Arbouet, C. Girard, R. Quidant, *Optics Express* **2010**, *18*, 3035.

[54] P. D. Terekhov, A. B. Evlyukhin, A. S. Shalin, A. Karabchevsky, *Journal of Applied Physics* **2019**, *125*, 173108.

[55] A. Alabastri, A. Toma, C. Liberale, M. Chirumamilla, A. Giugni, F. De Angelis, G. Das, E. Di Fabrizio, R. P. Zaccaria, *Optics Express* **2013**, *21*, 7538.

[56] A. V. Goncharenko, J. K. Wang, Y. C. Chang, *Physical Review B - Condensed Matter and Materials Physics* **2006**, *74*, 235442.

[57] S. Tuccio, L. Razzari, A. Alabastri, A. Toma, C. Liberale, F. De Angelis, P. Candeloro, G. Das, A. Giugni, E. D. Fabrizio, R. P. Zaccaria, *Optics Letters* **2014**, *39*, 571.

[58] S. Grafström, P. Schuller, J. Kowalski, R. Neumann, *Journal of Applied Physics* **1998**, *83*, 3453–3460.

[59] R. Tellez-Limon, M. Février, A. Apuzzo, R. Salas-Montiel, S. Blaize, *Journal of the Optical Society of America B* **2017**, *34*, 2147.

[60] B. Desiatov, I. Goykhman, M. Tzur, U. Levy, *Nano Letters* **2013**, 9–10.

[61] D. K. Gramotnev, S. I. Bozhevolnyi, *Nature Photonics* **2014**, *8*, 13–22.

[62] G. Palermo, U. Cataldi, A. Condello, R. Caputo, T. Bürgi, C. Umeton, A. De Luca, *Nanoscale* **2018**, *10*, 16556–16561.

[63] L. Pezzi, G. Palermo, A. Veltri, U. Cataldi, T. Bürgi, T. Ritacco, M. Giocondo, C. Umeton, A. D. Luca, *Journal of Physics D: Applied Physics* **2017**, *50*, 435302.

[64] G. E. Lio, G. Palermo, A. De Luca, R. Caputo, *Journal of Chemical Physics* **2019**, *151*, 244707.

[65] T. Miloh, *Physical Review Fluids* **2016**, *1*, 044105.

[66] K. W. Mauser, S. Kim, S. Mitrovic, D. Fleischman, R. Pala, K. C. Schwab, H. A. Atwater, *Nature Nanotechnology* **2017**, *12*, 770–775.

[67] D. Gall, *Journal of Applied Physics* **2016**, *119*, 085101.

[68] X. Y. Wang, D. M. Riffe, Y. S. Lee, M. C. Downer, *Physical Review B* **1994**, *50*, 8016–8019.